\documentclass[aps,pra,twocolumn,preprintnumbers,amsmath,amssymb,
superscriptaddress,floatfix,a4paper,longbibliography]{revtex4-2}%

\usepackage{graphicx}
\usepackage[dvipsnames]{xcolor}

\usepackage[utf8]{inputenc}
\usepackage{siunitx} %\SI
\usepackage[colorlinks,linkcolor=blue,citecolor=blue,urlcolor=blue]{hyperref}
\newcommand{\refsubfig}[2]{\hyperref[#1]{\ref*{#1}#2}}

\begin{document}

%TC:ignore

\title{Observation of non-adiabatic Landau-Zener tunneling among Floquet states}
%Floquet-Landau-Zener interferometry in experiment and theory

\author{Yun~Yen}
\affiliation{PSI Center for Scientific Computing, Theory, and Data, 
5232 Villigen PSI, Switzerland}
\affiliation{École Polytechnique Fédérale de Lausanne (EPFL), Switzerland}

\author{Marcel Reutzel}
\affiliation{I. Physikalisches Institut, Georg-August-Universit\"at G\"ottingen, G\"ottingen, Germany}

\author{Andi Li}
\affiliation{Department of Physics and Astronomy and Pittsburgh Quantum Institute, University of Pittsburgh, Pittsburgh, Pennsylvania, USA}

\author{Zehua Wang}
\affiliation{Department of Physics and Astronomy and Pittsburgh Quantum Institute, University of Pittsburgh, Pittsburgh, Pennsylvania, USA}

\author{Hrvoje Petek}
\email{petek@pitt.edu}
\affiliation{Department of Physics and Astronomy and Pittsburgh Quantum Institute, University of Pittsburgh, Pittsburgh, Pennsylvania, USA}
\affiliation{Department of Physics and Astronomy and IQ Initiative, University of Pittsburgh, Pittsburgh, Pennsylvania, USA}

\author{Michael~Sch\"uler}
\email{michael.schueler@psi.ch}
\affiliation{PSI Center for Scientific Computing, Theory, and Data, 
5232 Villigen PSI, Switzerland}
\affiliation{Department of Physics, University of Fribourg, CH-1700 Fribourg, Switzerland}

\begin{abstract}

% Electromagnetic fields not only drive electronic transitions but also create light-matter coupled phases beyond equilibrium. As well established, Floquet engineering is a paradigmatic method to dress electronic structure with time periodic fields, forming quasi-stationary Floquet states. With increasing field strength, interesting non-perturbative responses of the dressed states emerge, yet their nonlinear dynamics remain challenging to interpret.  With a joint experiment-theory approach, here we discuss the photoemission signature of Floquet states on Cu(111) surface across different light-matter interaction regimes, using interferometrically time-resolved multi-photon photoemission spectroscopy (ITR-mPP) - a nonlinear coherent variant of angle-resolved photoemission spectroscopy. Remarkably,  at the highest  light field strengths, new photoemission feature appears, which we attribute to non-adiabatic Landau-Zener tunneling among Floquet states.  We adapt the instantaneous Floquet state (IFS) formalism to examine real-time excitation pathways in non-perturbative photoemission, and emphasize the direct connection between the onset of light-dressing of matter, the non-perturbative ultrafast lightwave electronics, and the  high-optical-harmonic generation in the solid-state. 

Electromagnetic fields not only induce electronic transitions but also fundamentally modify the quantum states of matter through strong light-matter interactions. As one established route, Floquet engineering provides a powerful framework to dress electronic states with time-periodic fields, giving rise to quasi-stationary Floquet states. With increasing field strength, non-perturbative responses of the dressed states emerge, yet their nonlinear dynamics remain challenging to interpret. In this work we explore the emergence of non-adiabatic Landau-Zener transitions among Floquet states in Cu(111) under intense optical fields. At increasing field strength, we observe a transition from perturbative dressing to a regime where Floquet states undergo non-adiabatic tunneling, revealing a breakdown of adiabatic Floquet evolution. These insights are obtained through interferometrically time-resolved multi-photon photoemission spectroscopy, which serves as a sensitive probe of transient Floquet state dynamics. Numerical simulations and the theory of instantaneous Floquet states allow us to directly examine real-time excitation pathways in this non-perturbative photoemission regime. Our results establish a direct connection the onset of light-dressing of matter, non-perturbative ultrafast lightwave electronics, and high-optical-harmonic generation in the solids. 

\end{abstract}

\maketitle
\noindent 

%TC:endignore

% Outline

\section*{Introduction}

%The study of high optical harmonic generation and the related research in attosecond electron physics motivated research on the light driven nonperturbative light--atom and --molecules interactions~\cite{corkum_attosecond_2007,hentschel_attosecond_2001,noauthor_attosecond_nodate}. In Keldysh theory, the Keldysh parameter distinguishes the perturbative quantum response for $\gamma \propto \omega (2I_0)^{1/2}/E_0  \gg 1$ from the nonperturbative tunnel ionization regime for $\gamma \ll 1$~\cite{perelomov_ionization_1966,zheltikov_keldysh_2016}, where $\omega$ is the optical circular frequency, $I_0$, the state ionization potential, and $E_0$, the electric field strength in vacuum.

The study of high optical harmonic generation and the related research in attosecond electron physics motivated research on the light driven nonperturbative light-matter interaction~\cite{corkum_attosecond_2007,hentschel_attosecond_2001,hommelhoff_attosecond_2015-1}. The onset of non-perturbative responses of solids to optical fields is subtle due to their periodic crystalline structures and the dielectric screening. When applied to semiconductors and insulators, valence electrons can be promoted to and accelerated into conduction bands, with the possibility of rescattering with holes, leading to high harmonic generation. Such strong field driven phenomena are central to the burgeoning field of lightwave electronics where THz fields drive PHz frequency electronic responses ~\cite{ghimire_high-harmonic_2019,ghimire_observation_2011,hohenleutner_real-time_2015,langer_lightwave-driven_2016,dienstbier_tracing_2023}. The negative real part of the dielectric functions of metals, $\mathrm{Re}(\epsilon)<0$~\cite{yang_optical_2015}, however, entangles the optical fields ~\cite{liebsch_dynamical_1987,apell_non-local_1984} with their free electron screening responses leading to non-local field intensification on attosecond time scales that is experienced as familiar mirror reflection. The enhanced surface fields can drive perturbative multiphoton quantum transitions evident in one or two color multiphoton photoemission (mPP) processes~\cite{reutzel_coherent_2020,reutzel_coherent_2019,li_multidimensional_2022,reutzel_nonlinear_2019,dreher_focused_2023,beyazit_ultrafast_2023}. As the driven surface field strength approaches the level of electronic Coulomb potentials, the electronic structure is dressed on an attosecond time scale, offering novel approaches to tailor the electronic structure of matter with light. 

Floquet engineering -- a paradigm for light-matter dressing with periodic fields -- can be used to control the electronic band structure, including its symmetry and band topology~\cite{oka_photovoltaic_2009,oka_floquet_2019,rudner_band_2020,mciver_light-induced_2020,hubener_creating_2017,vu_light-induced_2004}. In common  applications, low frequency fields generate Floquet side bands within band gaps of semiconductors and insulators~\cite{wang_observation_2013,zhou_pseudospin-selective_2023,ito_build-up_2023,mahmood_selective_2016,aeschlimann_survival_2021}, where they have been mapped in the energy-momentum space by angle-resolved photoemission spectroscopy~\cite{wang_observation_2013,mahmood_selective_2016,merboldt_observation_2024,choi_direct_2024}. Resonant matter polarization fields, such as excitons ~\cite{kobayashi_floquet_2023,pareek_driving_2024,chan_giant_2023} or plasmons~\cite{gumhalter_electron_2022} can also spectroscopically impose multi-quanta dressed Floquet states. It is thus important to study the field-induced build-up time and the Floquet states dynamics~\cite{ito_build-up_2023,schuler_how_2020,aeschlimann_survival_2021} on optical cycle time scales.

\begin{figure*}[t]
    \includegraphics[width=1.0\textwidth]{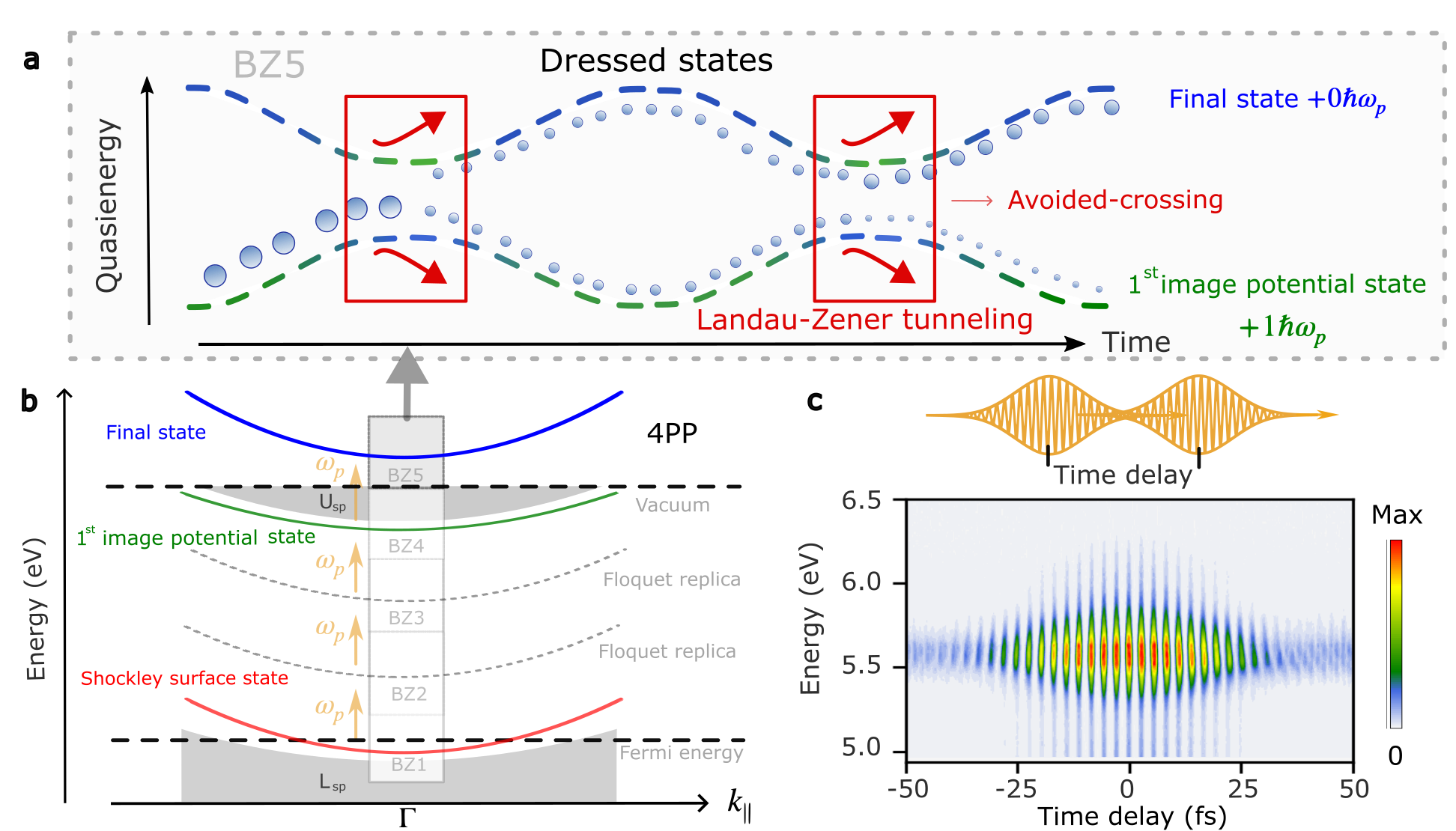}
    \caption{\textbf{Schematic of interferometric multiphoton photoemission process in strong field regime.} \textbf{a } Landau-Zener (LZ) mechanism among dressed states in photoemission. Dressed photoelectron final states form avoided crossings (AC) with other Floquet states, where the dressed final states gain electron population via LZ tunneling. \textbf{b} Sketch of the band dispersion of the Shockley surface state (SS), first image potential state (IP1), a photoelectron final state, and their Floquet replicas in different Floquet Brillouin zones (BZ). Electrons are photoemitted from the occupied SS through IP1 into the final state continuum, leading to 4-photon photoemission (4PP). SS and IP1 lie within the surface projected band gap between lower and upper \textit{sp}-band $L_{sp}$ and $U_{sp}$ (gray shading). $\omega_p$ denotes the photon energy. \textbf{c} An example of the experimental interferometric-time-resolved 4PP data at $\Gamma$ with field strength 2.2~V/nm. Two identical pulses dress and probe Cu(111) as the phase delay defines strength of the driving field. The energy axis is defined with respect to Fermi energy.}
    \label{fig:schem_exp}
\end{figure*}

%Projected band gaps in metals, where the surface states are decoupled in momentum and energy from the continuous bulk bands, offer a context to study the Floquet physics of nearly-ideal two-level, free electron-like bands. 

Here, we investigate the coherent dressing of Cu(111) surface states, which offers an ideal two-level platform to explore high-field responses in the presence of many-body screening effects~\cite{frisk_kockum_ultrastrong_2019}.  To this end, we employ coherent interferometrically time-resolved multi-photon photoemission spectroscopy (ITR-mPP).
At lower field strengths, this method has been used to study Floquet side bands, where their Autler-Townes (AT) splitting becomes visible in the spectra~\cite{reutzel_coherent_2019,reutzel_coherent_2020}. We extend this study to higher fields to explore the transition from the perturbative to nonperturbative responses with attosecond phase precision and field  amplitude control. We show from experimental mPP data that the onset of additional structure at high field strengths can no longer be described within the Floquet dressing picture, but stands as a signature of non-adiabatic dynamics -- particularly Landau-Zener (LZ) tunneling among the dressed states, as illustrated in Fig.~\ref{fig:schem_exp}\textbf{a}. We establish that such non-adiabatic tunneling among Floquet states is a general feature of the non-perturbative response in strong-field nonlinear light-matter interactions.

%We establish a clear physical picture of the LZ tunneling as a general feature emerging in the nonperturbative response within the basis of Floquet states in high field nonlinear light-matter interactions. 

\section*{Results}

\subsection{Experimental results}%Dressed surface electronic structure and Coherent mPP on Cu(111)}
% Surface band structure & experimental details
% Try with 
The discontinuity of a periodic crystal lattice at metal-vacuum interfaces supports partially occupied Shockley surface states (SS). Similarly, the Coulomb interaction between a vacuum electron and its screening image charge forms a Rydberg-like series of image potential states (IP)% whose energy spectrum asymptotically converges to the vacuum level
~\cite{chulkov_image_1999,chulkov_electronic_2006}. We explore imprinting of the non-adiabatic surface response in interferometric nonlinear four-photon photoemission (4PP) electron distributions when the photon energy is tuned to the three-photon resonance from the SS to the first image potential state (IP1) at $\Gamma$ ($k_{\parallel}=0$). Because the decoupling of surface states from the bulk bands and the coherent two-level optical response of Cu(111) are well-documented by ultrafast photoemision spectroscopy~\cite{Aeschlimann25ss}, it is an ideal platform for studying the onset of non-perturbative light-matter interaction with momentum-resolution. A sketch of the surface electronic structure is demonstrated in Fig.~\ref{fig:schem_exp}\textbf{b}.% for optical dressing by intense light pulses. 

To study the dressing of the surface electronic structure, two identical 20~fs laser pulses excite the ITR-4PP signal by scanning their time delay with 50 attosecond precision (Fig.~\ref{fig:schem_exp}\textbf{c}). The photon energy $\omega_p = 1.54$~eV is set to one-third of the gap between SS and IP1 at the $\Gamma$ point to enhance the four-photon excitation above the vacuum level by the IP1-SS three-photon resonance. The optical field is sufficiently intense to excite five photon above threshold photoemission (ATP) 5PP signal~\cite{reutzel_above-threshold_2020}. A schematic of the mPP process is shown in Fig.~\ref{fig:schem_exp}\textbf{b}. In ref. ~\cite{reutzel_coherent_2020}, it is shown that as the laser field strength is strong enough, ITR-4PP spectra of Cu(111) reveal formation of a Floquet engineered band structure that is characterized by a momentum dependent AT splitting of the coupled bands%exhibit onset of the Floquet splitting arising from strong-field Stark effect as the optical Rabi frequency approaches the energy gap between SS and IP1
. Fig.~\ref{fig:schem_exp}\textbf{c} shows the ITR-4PP interferogram at $\Gamma$ point with a moderate estimated experimental field strength of $F_0^{\mathrm{exp}} \sim$ 2.2 V/nm. The signal shows nonlinear 4PP oscillation at the driving frequency($\omega_p$), with an oscillation period of 2.7 fs. The decay of the IP1-SS coherence is much slower than the laser pulse~\cite{ogawa_optical_1997,knoesel_temperature_1998}. Therefore, the dressing
of the surface states remains coherent and robust up to a large time delay ($\sim 10$ fs). %The field strength dependent ITR-4PP is shown in Fig.~\ref{fig:tsurff_exp}\textbf{e}-\textbf{g}.

%In the present work, 
Additional ITR-4PP interferograms at $\Gamma$ recorded for increasing field strength is shown in Fig.~\ref{fig:tsurff_exp}\textbf{e}-\textbf{g}. Fourier analysis of the ITR-4PP interferograms reveals the harmonic content of the coherent nonlinear polarization \cite{reutzel_coherent_2020,reutzel_coherent_2019,li_unraveling_2013}; we show the energy-, time-, and field-strength-resolved amplitudes of the non-resonant 2$\omega_p$ coherent polarization field in Fig.~\ref{fig:tsurff_exp_2w}\textbf{a,b,d,e} %We extract the pulse delay-dependent dressed two-level electronic structure by performing inverse Fourier transform of the signal Fourier component at oscillation frequency $2\omega_p$ 
(See Methods for experimental details). The amplitude of the 2$\omega_p$ harmonic signal at a field strength of $F_0^{\mathrm{exp}} \sim$ 2.2 V/nm in Fig.~\ref{fig:tsurff_exp_2w}\textbf{a} shows a clear two-fold splitting of the Fourier amplitude, which  we interpret as the AT splitting of the light-matter coupled Floquet states as the joint pulse field strength reaches its maximum at zero delay~\cite{reutzel_coherent_2020}. The interest here is when the optical field strength is further increased (Fig.~\ref{fig:tsurff_exp_2w}\textbf{d}), the pulse delay dependent spectrum deviates from the two-fold splitting and acquires a dominant central peak between the AT dressed bands. Examining the phase of the 2$\omega_p$, the low field AT split responses oscillate in-phase with each other and the driving field, but the high field response shows a distinct $\pi$ phase shift relative to the AT side peaks (Fig.~\ref{fig:tsurff_exp_2w}\textbf{e}).  

% We interpret the $\pi$ phase shifted response of the central peak peak as the  signature of the nonperturbative \textcolor{red}{attosecond} light-matter interaction beyond the quasistationary Floquet dressing.%, because the number of Floquet side bands for a two level system is maximally two. 

%with three different field strength displayed in fig.\ref{fig:schem_exp}\textbf{b}-\textbf{i}. The splitted 4PP in fig.\ref{fig:schem_exp}\textbf{d} and \textbf{e} is consistent with the Floquet splitting picture[], where the two-fold splitting is derived from SS and IP1. In fig.\ref{fig:schem_exp} \textbf{f}-\textbf{g}, a salient central peak shows up with a peculiar $\pi$ phase shift relative to the other two side bands. This central peak can not be identified as any Floquet state. In such strong field regime, the mPP signal consist of highly non-perturbative contribution beyond probing of the light-dressed states, and its relation with Floquet state dynamics is our main result here.

\subsection{Non-equilibirum Green's functions simulation}
To simulate the field strength dependent ITR-mPP results, we calculate photocurrent within time-dependent non-equilibrium Green's function (td-NEGF) formalism. We use a one dimensional real-space model Hamiltonian with Chulkov-type potential $\hat{V}_c(z)$~\cite{chulkov_image_1999,chulkov_electronic_2006} (See supplementary material~\cite{supplement}), in order to reproduce the surface state eigenenergies and band gaps. The light-matter interaction is described with the velocity gauge minimal coupling as
\begin{align}
    \label{eq:tsurff_ham}
    \hat{H}(z,t) = \frac{1}{2}\left[-i \frac{\partial}{\partial z}  - A(t)\right]^2+\hat{V}_c(z) \ ,
\end{align}
where the vector potential $A(t)$ describes the perpendicular effective field at the surface. This formalism captures non-perturbative responses including ATP and laser-assisted photoemission (LAPE)~\cite{saathoff_laser-assisted_2008,keunecke_electromagnetic_2020} (See Methods for details of the simulations).

% The central quantity for calculating the photoemission intensity is the one-body lesser Green's function $G^<_{ij}(t, t')$, which is represented in the basis of the SS and IP1 states ($i,j =$SS, IP1). The Green's function is obtained by solving the Kadanoff-Baym equations (See Methods for details). The td-NEGF approach allows to explicitly include the finite lifetime of the IP1 state~\cite{neppl_direct_2015} as well as the non-perturbative photoemission through a self-energy. Adopting the gauge-invariant theory for time-resolved photoemission from ref.~\cite{schuler_theory_2021}, the intensity is calculated from
% \begin{align}
%     \label{eq:meier-wingreen}
%     I(E) &\propto  \mathrm{Re} \sum_{ij} \int_0^{\infty} dt \int^t_0 dt' \Sigma^{>}_{ij}(E;t,t') G^{<}_{ji}(t',t) \ .
% \end{align}
% Here, $\Sigma^{>}_{ij}(E;t,t') = M^*_{i}(E) g^>(E, t, t') M_{j}(E)$ is the self-energy describing the continuum of photoelectron states, including the photoemission matrix elements $M_i(E)$ (calculated from the Hamiltonian~\eqref{eq:tsurff_ham}) and the Green's function $g^>(E, t, t')$ of the photoelectrons in the presence of the field. 
%This formalism captures non-perturbative responses including ATP and laser-assisted photoemission (LAPE)~\cite{saathoff_laser-assisted_2008,keunecke_electromagnetic_2020}.

\begin{figure*}[t]
    \includegraphics[width=1.0\textwidth]{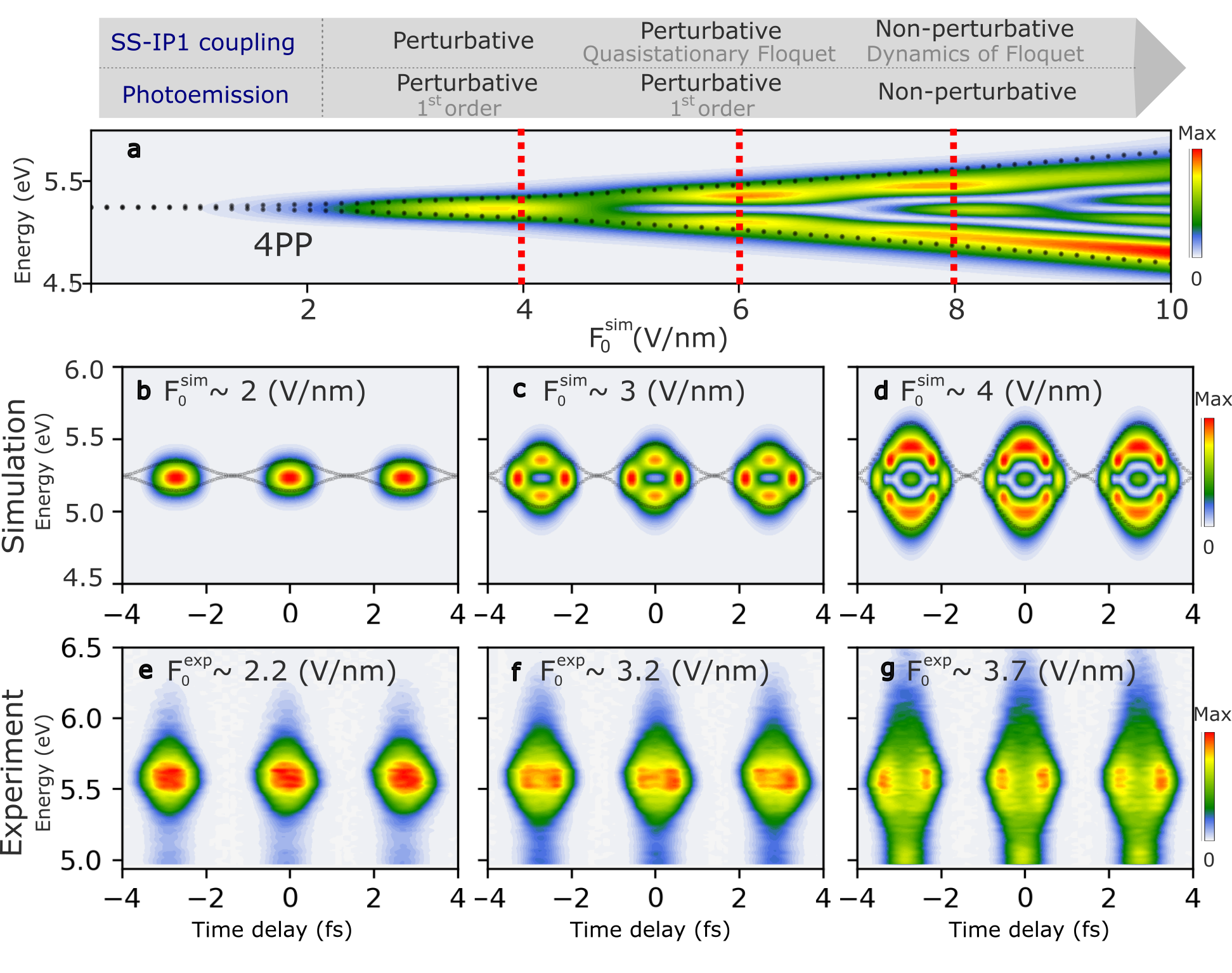}
    \caption{\textbf{Field strength dependent NEGF simulation and experiment.} \textbf{a} Simulated field dependent spectrum (color coding) and the calculated Floquet quasienergy (black dots). The Floquet quasienergies are calculated with basis of SS and IP1, using a continuous sine wave field (See Methods), which can be justified if the Gaussian pulse is sufficiently long. \textbf{b}-\textbf{d} ITR-4PP NEGF simulation with peak field strength $F_0^{\mathrm{sim}} \sim$ 2, 3, and 4 V/nm and the corresponding Floquet quasienergies calculated with effective field strength averaging over two pulses (black dots). \textbf{e}-\textbf{f} Experimental ITR-4PP data with estimated field strength $F_0^{\mathrm{exp}} \sim$ 2.2, 3.2, and 3.7 V/nm. Here we use $F_0^{\mathrm{sim}}$ and $F_0^{\mathrm{exp}}$ to denote theoretical and experimental field strength. The energy axis is the referenced to the Fermi energy, with a theoretical work function of 4.5 eV in the simulation. }
    \label{fig:tsurff_exp}
\end{figure*}

\begin{figure*}[t]
    \includegraphics[width=0.9\textwidth]{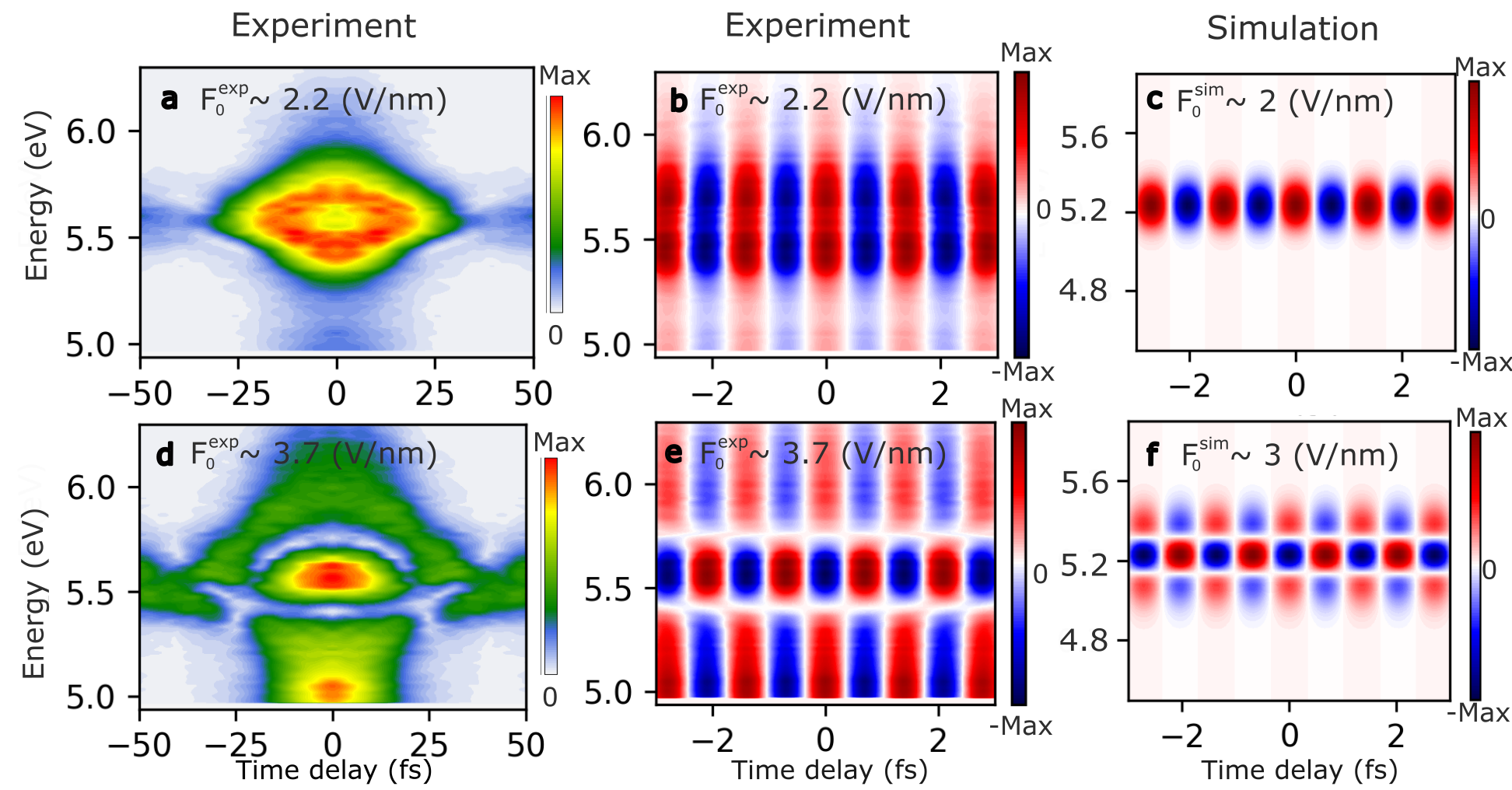}
    \caption{ \textbf{Comparison of the 2$\omega_p$ coherent polarization fields in experiment and simulation.} The 2$\omega_p$ harmonics is obtained by moving window forward/backward Fourier transformation and inverse Fourier transformation with a Gaussian profile centered at 2$\omega_p$. \textbf{a} The envelope function and the \textbf{b} 2$\omega_p$ oscillation at $F_0^{\mathrm{exp}} \sim 2.2$ V/nm shows two-fold in phase splitting. \textbf{d} The envelope function and the the \textbf{e} 2$\omega_p$ oscillation at $F_0^{\mathrm{exp}} \sim 3.7$ V/nm shows three-fold splitting with a $\pi$ phase shift in the central peak. \textbf{c,f} The simulated 2$\omega_p$ oscillation at $F_0^{\mathrm{sim}} \sim 2$ V/nm and $F_0^{\mathrm{sim}} \sim 3$ V/nm. The energy axis is the referenced to the Fermi energy, with a theoretical work function of 4.5 eV in the simulation.} 
    \label{fig:tsurff_exp_2w}
\end{figure*}

To first illustrate the field-strength dependence of 4PP, we simulate spectrum with a single pulse, as a function of Gaussian peak field strength $F_0^{\mathrm{sim}}$ in Fig.~\ref{fig:tsurff_exp}\textbf{a}. The 4PP spectrum develops observable two-fold energy splitting structure at a moderate field strength of $F_0^{\mathrm{sim}} \sim 5$~V/nm and acquires additional structure as field strength increases. Up to fields of $F_0^{\mathrm{sim}} \sim 7 $~V/nm, the 4PP primarily probes the AT splitting of Floquet states quasienergies, as represented with the black dots in Fig.~\ref{fig:tsurff_exp}\textbf{a}. As $F_0^{\mathrm{sim}}$ gets stronger, distinct central peaks start to emerge in a spectral range away from the Floquet quasienergies. In Fig.~\ref{fig:tsurff_exp}\textbf{b}-\textbf{d}, we select three different peak field strengths $F_0^{\mathrm{sim}}$ (red dashed lines in Fig.~\ref{fig:tsurff_exp}\textbf{a}), to simulate different regimes of the corresponding ITR-4PP interferograms. The field dependence demonstrates the evolution from two-fold to multi-fold splitting, in accordance to the behavior observed in the experiments. 
The effective Floquet quasienergies in Fig. ~\ref{fig:tsurff_exp}\textbf{b}-\textbf{d} roughly follows the boundary of ITR-4PP. At comparable field strengths, the NEGF simulation reproduces our main experimental finding: the emergence of the central  $\pi$ phase shift in the 2$\omega_p$ component of ITR-4PP data shown in Fig.~\ref{fig:tsurff_exp_2w}\textbf{c} and \textbf{f}.

From the comparison of the splitting structure and Floquet quasienergies in Fig.~\ref{fig:tsurff_exp}\textbf{a-d}, we identify three distinct regimes of the photoemission: (i) For weak fields $F_0^{\mathrm{sim}} < 5$~V/nm, the 4PP spectrum follows  SS to IP1 resonant excitation under the Fermi's golden rule. (ii) Intermediate field strengths $ 5$ V/nm $< F_0^{\mathrm{sim}} < 7$ V/nm induce significant perturbative dressing of SS and IP1, giving rise to two-fold split Floquet states. (iii) In the strong-field regime $F_0^{\mathrm{sim}} > 7$~V/nm, however, a multifold splitting of bands appears that cannot alone be explained by Floquet physics. Clearly, at high field strengths, nonlinear responses emerge beyond probing of the quasi-stationary light-dressed states.

%To see this, assuming the pulse profile is infinitely long, time-resolved ARPES intensity can be written in a linear response fashion as \cite{sentef2013examining, freericks2009theoretical}
%\begin{align}
%    \label{eq:ness}
%    I(E,t,t') \propto Im\sum_{jj'} &\int_0^{\infty} \int_0^{\infty} dt dt' M^{*}_j(p) \nonumber\\
%    &\times G^{<}_{jj'}(t,t')M_{j'}(p)e^{-i\Phi(t,t')} 
%\end{align}
%, where $\Phi(t,t')=\int_{t}^{t'} d\tau [\omega_p - \epsilon_p(\tau)]$, $\epsilon_p$ is energy for final state of momentum $p$ with energy conservation $E=p^2/2$, and matrix element $M_j(E)$ couples band $j$ to the final states. The physical lesser Green's function $G^{<}_{jj'}$ can be recovered from lesser Floquet Green's function within non-equilibrium steady state (NESS) formalism \cite{schuler2020circular}, which contains the dynamically equilibrated spectral information of Floquet states. As a result, Floquet states are perturbatively probed in photoemission.  %However, in the strongest field regime $F_0 > 80 (MV/cm)$, Floquet NESS Green's function fails to describe the emerging multifold splitting peaks. Clearly significant nonlinear responses beyond probing of the quasi-stationary light-dressed states contribute to 4PP.

% Some description of the regime...

%\section*{Instantneous Floquet states (IFS) formalism}
\subsection{Instantaneous Floquet excitation pathway}

%We now provide a physical interpretation for the time dependent evolution of the dressed eigenstates by considering how the dynamics of Floquet states give rise to the 4PP spectra. Whereas the Floquet calculation in Fig.~\ref{fig:tsurff_exp}\textbf{a} relies on the assumption of the optical field time periodicity, the field driving strength can cause the dynamics among Floquet states to become non-perturbative. In this regime, the photoemission process can be interpreted based on the real-time Floquet state population dynamics on sub-cycle time scale.

%The Floquet calculation in Fig.~\ref{fig:tsurff_exp}\textbf{a} relies on the assumption of the field time periodicity. However, strong field intensity is possible to induce non-adiabatic dynamics among Floquet states, which is missing in the quasistationary Floquet picture. In order to describe the photoemission with the basis of Floquet states, it is necessary to incorporate the time dependence of Floquet quasienergies and wavefucntion. This would provide a clear physical interpretation for the strong field process, as we find that the peculiar 4PP splitting branches can be interpreted with the sub-cycle nonadiabatic population dynamics of Floquet states.

% New (direction first)
The agreement between experiments and td-NEGF simulation indicates that the photoemission dynamics can be captured within the manifold of SS, IP1, and photoelectron final states. Nevertheless, the physical origin of the strong field multifold splitting is still unclear.

Now we investigate how the dynamics of Floquet states produce the additional splitting features and the $\pi$ phase shift. To get a physical picture of the mPP process, we need to interpret the excitation pathways in terms of Floquet states, and understand \textit{how electrons are promoted from the initial SS via Floquet states to the photoemission final states in real time}. This population dynamics in strongly driven systems can be understood in the language of instantaneous Floquet states (IFS). For a single pulse with sufficiently many cycles, following the envelop of the pulse, we can define Floquet states at the instantaneous field strength. For a sufficiently strong instantaneous field, the Floquet states form avoided crossings that lead to non-adiabatic dynamics~\cite{ikeda_floquet-landau-zener_2022}. As we will show, the extra 4PP branches are signatures of such non-adiabatic dynamics, as shown in the sketch in Fig.~\ref{fig:schem_exp}\textbf{a}.

%Nevertheless, as the driving strength increases such that the generalized Rabi frequency approaches $\omega_p$, coupling among the Floquet ladder states can lead to nontrivial dynamics. In this regime, the photoemission process can be described by the real-time Floquet state population dynamics on sub-cycle time scale.

Any driven state $|\Phi(t_{r})\rangle$ can be expanded with a set of IFS at time $t_{r}$ as $| \Phi(t_{r}) \rangle = \sum_{\alpha m} c_{\alpha m}(t_{r}) | \phi_{\alpha,m}(t_{r}) \rangle$, where $|\phi_{\alpha,m}(t_{r})\rangle$ is the m$^{\mathrm{th}}$ mode of IFS $\alpha$ (See Methods).
Here the photoemission dynamics is modeled with a set of IFS originating from the three relevant states : $\alpha =$ SS, IP1, and a photoelectron final state with energy $E_{\mathrm{final}}$. The Floquet states excitation pathway is described by the population dynamics $|c_{\alpha m}(t_{r})|^2$ of the IFS, following interaction with two identical Gaussian pulses with a phase (time) delay. We take the final states to be time-reversed low-energy electron diffraction (LEED) states (See supplementary material~\cite{supplement}). The mPP signal at energy $E_{final}$ is given by  the corresponding final state population $\sum_m |c_{\mathrm{final},m}(t_r = t_{\mathrm{end}})|^2$ at the end of the pulses. The IFS can form avoided crossings, where the non-adiabatic dynamics can be described by LZ tunneling.

%\textcolor{red}{
%[Here you need to explain the key advantage of the IFS formalism. Instead of solving the TDSE in terms of the Floquet states, we separate the evolution into adiabatic evolution and LZ transitions in the between the Floquet states].
%}

\begin{figure*}[t]
    \includegraphics[width=1.0\textwidth]{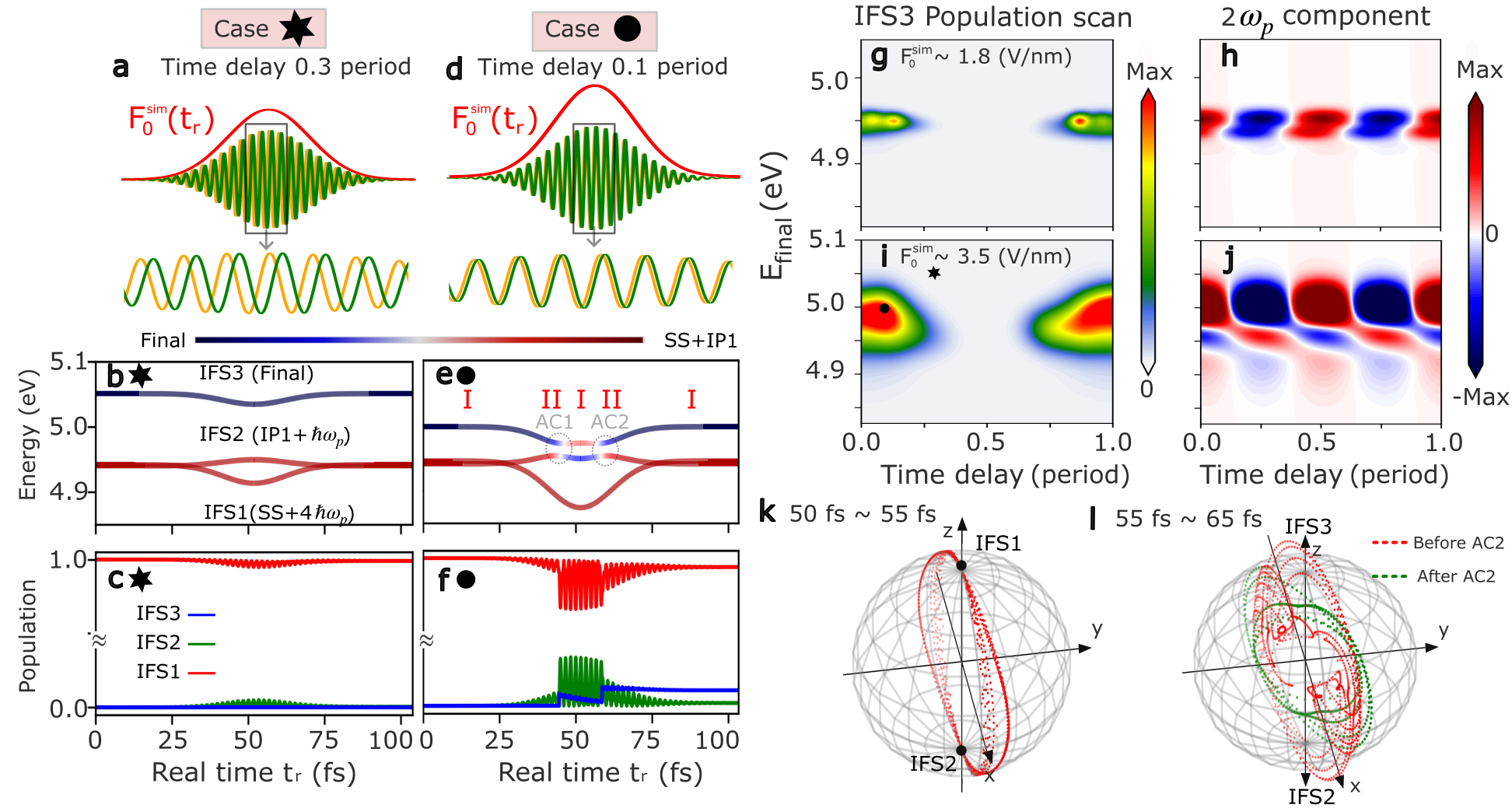}
    \caption{\textbf{Dynamics of instantaneous Floquet states.} \textbf{a-f} Two example cases of the IFS dynamics. Two identical Gaussian pulses with time delay \textbf{a} 0.3 period and \textbf{d} 0.1 period are used to calculate the corresponding real-time ($t_r$) dependent IFS quasienergy spectrum with final state energy \textbf{b} $E_{\mathrm{final}} \sim 5.05$ eV and \textbf{e} $E_{\mathrm{final}} \sim 5.0$ eV. The color-coding shows the character of each IFS, i.e. projection of each IFS onto three relevant states (SS,IP1, and final state).  \textbf{c,f} The corresponding time dependent population of three IFS, where the population of IFS3 is multiplied by 5 for visualization. IFS2 and IFS3 form avoided crossing points (AC1, AC2) in \textbf{e}, and the Landau-Zener mechanism leads the population jump in \textbf{f}. \textbf{g,i} Population of IFS3 at the end of the pulse with varying final state energy ($E_{\mathrm{final}}$) and time delay, where the shaded circle and star in \textbf{i} correspond to two cases in \textbf{a-f}. \textbf{h,j} The corresponding 2$\omega_p$ harmonics. \textbf{k,l} Coherent oscillation of the driven wavefunction in the case circle calculation (\textbf{d}-\textbf{f}) visualized with Bloch sphere for \textbf{k} IFS1 and IFS2 between $50$ fs and $55$ fs and \textbf{l} IFS2 and IFS3 between $55$ fs and $65$ fs. Radius for both sphere is normalized for better visualization. The energy axes and $E_{\mathrm{final}}$ are referenced to the theoretical Fermi energy with a theoretical work function of 4.5 eV. }
    
    %the norm of two relevant states at 40 fs for \textbf{m} and 60 fs for \textbf{n}. Therefore all IFS fall on the z axis but not on the poles of the spheres. } 
    % a-c : put two cases with different delta t, and draw the corresponding star in d,g..., we need to stress the contexts more with two pulses, describe 'why' context  in the texts, change a(t0) to F(t0) and plot them on top of the graphs...!!
    \label{fig:ifs}
\end{figure*}

To illustrate, we show two distinct cases of the IFS dynamics leading to photoemission labeled by a \textit{star}  (Fig.~\ref{fig:ifs}\textbf{a-c}) and a \textit{circle} (Fig.~\ref{fig:ifs}\textbf{d-f}). They are simulated with different final state energies $E_{\mathrm{final}}$ as the time delay between two Gaussian pulses tunes the states to possible avoided crossings where the LZ mechanism occurs (\textit{circle}) or not (\textit{star}). For the \textit{star} (\textit{circle}) case, the Gaussian pulses with time delay 0.3 period (0.1 period) are illustrated in Fig.~\ref{fig:ifs}\textbf{a} (\textbf{d}). The different time delays define delay dependent frequencies and amplitudes. The instantaneous photon energy is set to be constant $\omega(t_r)=\omega_p$, and $F_0^{\mathrm{sim}}(t_r)$ is the instantaneous field strength following the envelope for the addition of two pulses. The final states have energy $E_{\mathrm{final}} \sim 5.05$~eV for the \textit{star} case and $E_{\mathrm{final}} \sim 5.0$~eV for the \textit{circle} case, respectively.

Fig.~\ref{fig:ifs}\textbf{b,e} show the corresponding real time ($t_r$) dependent IFS quasienergy spectrum within one Floquet Brillouin zone, with the color coding representing the projection of each IFS onto the three relevant states. At the beginning of the excitation ($ t_r=0$~fs), IFS1, IFS2, and IFS3 are the exact copies of SS, IP1 and the final state, because mixing among them has yet to occur, and IFS1 and IFS2 are degenerate under three photon resonant condition. As the field strength gradually increases following the pulse envelope overlap, the splitting of IFS1 and IFS2 ensues. At $ t_r \sim 50 fs$, the peak field strength causes IFS3 to acquire some SS/IP1 character. %Simultaneously, IFS3 with initial energy $E_f \sim 0.5$~eV gradually deviates in energy by acquiring the AC Stark shift. 

In the \textit{star} case, the population dynamics is entirely simulated with a finite-space time dependent Schr{\"o}dinger equation (TDSE)(See Methods). Whereas in the \textit{circle} case, IFS2 and IFS3 form avoided crossings at time $t_r \sim 40$~fs (AC1) and time $t_r \sim 60$~fs (AC2). We categorize the dynamics into type \textit{I} and \textit{II}, as labeled in Fig.~\ref{fig:ifs}\textbf{e}. For type \textit{I}, the dynamics are simulated with a finite-space TDSE as in the \textit{star} case. At the beginning of the pulse, only IFS1 is populated since SS is the only occupied state. The time dependent populations (Fig.~\ref{fig:ifs}\textbf{c,f}) show clear coherent oscillation between IFS1 and IFS2 with the amplitude proportional to the field envelope. The contribution of coherent oscillation to IFS3 population is minimal,  because the matrix elements between the final state and two surface states are much smaller than the matrix element between SS and IP1.  Consequently, the population dynamics of IFS3 in type \textit{I} region remains adiabatic. The time-dependent wave-function coefficient $c_{\alpha m}(t_r)$ projected onto the manifold of IFS1 and IFS2 can be visualized on a Bloch sphere in Fig.~\ref{fig:ifs}\textbf{k}, where the wave-function trajectory evolves with a constant radius, indicating that the coherent populations dynamics evolve mostly between the IFS1 and IFS2.

In the \textit{circle} case, however, upon crossing AC1, the type \textit{II} dynamics becomes non-adiabatic and is described by a LZ transfer matrix~\cite{ikeda_floquet-landau-zener_2022}
\begin{equation}
    \label{eq:lz_matrix}
    \hat{\mathbf{T}}_{ii'} = 
    \begin{pmatrix}
        \sqrt{1-P_{ii'}}e^{-i\phi^{s}_{ii'}} & -\sqrt{P_{ii'}} \\
        \sqrt{P_{ii'}} & \sqrt{1-P_{ii'}}e^{+i\phi^{s}_{ii'}} 
    \end{pmatrix} \ ,
\end{equation}
where $i=\{\alpha m\}$ denotes the $m^{th}$ Floquet replica of IFS $\alpha$. The Stokes phase $\phi^{s}_{ii'}$ and $P_{ii'}$ are defined by details of the avoided crossing gaps (See Methods). Due to the finite off-diagonal elements in Eq.~\eqref{eq:lz_matrix}, part of the population exchange between IFS2 and IFS3 at both AC1 and AC2. On the Bloch sphere formed within the space of IFS2 and IFS3 in Fig.~\ref{fig:ifs}\textbf{l}, the Landau-Zener mechanism significantly alters the orbits before and after AC2. Because the dynamics of IFS3 remains adiabaic in the type-\textit{I} part, after crossing AC2,  most of the population at IFS3 persist until the end of the pulse, leading to a finite photoemission signal. In contrast, the absence of LZ tunneling in the \textit{star} case does not enhance the 4PP signal. In this picture, we are able to demonstrate how exactly the mPP process occurs in real time: the population oscillates between SS and the intermediate IP1 coherently, and "\textit{jumps}" to the final state via a non-adiabatic LZ mechanism.

To elucidate the role of LZ tunneling, we simulate the interferometic 4PP by recording the polulation of IFS3 for the entire range of $E_{\mathrm{final}}$ and varying time delay between the two pulses over one cycle. The simulation in Fig.~\ref{fig:ifs}\textbf{g}-\textbf{j} is performed for two peak field strengths ($F^{\mathrm{sim}}_0 = 1.8$~V/nm and $F^{\mathrm{sim}}_0=3.5$~V/nm), in order to show the impact of the non-adiabatic dynamics on ITR-4PP. The IFS approach remarkably reproduces the field strength dependent experimental 4PP features - the emergence $\pi$ phase shift in strong field in Fig.~\ref{fig:ifs}\textbf{j}. The $\pi$ phase shift naturally arises from the distinct triangular shape of the ITR-4PP in the $E_{\mathrm{final}}$-time delay plane, as shown in Fig.~\ref{fig:ifs}\textbf{g} and \textbf{i}. The boundary of finite 4PP intensity region is approximately outlined by the "\textit{touching}" points between IFS3 and IFS1/IFS2 (See supplementary material~\cite{supplement}). This aligns with our simulation in Fig.~\ref{fig:tsurff_exp}\textbf{b}-\textbf{d}, where the effective Floquet quasienergies delineate the border line of finite ITR-4PP. A stronger field strength enhances Floquet quasienergy splitting and thereby broadens the range of 4PP. For instance, Fig.~\ref{fig:ifs}\textbf{i} shows enhanced photoemission signal between the in-phase fringe maxima compared to Fig.~\ref{fig:ifs}\textbf{g}. As a result, the ITR-4PP takes on a triangular shape and the corresponding 2$\omega_p$ component shows a $\pi $ phase shift at the center of the 4PP (See supplementary material~\cite{supplement} for a more detailed discussion.).

\section*{Discussion}

We have systematically explored different regimes of light-matter interaction on the Cu(111) surface. For moderate field strength, the 4PP spectrum shows evidence of Floquet splitting as the field strength increases~\cite{reutzel_coherent_2020}, which is explained with light dressing of the SS-IP1 two level system. Identifying such light-induced hybridization phenomena in conventional pump-probe optical and photoemission experiments has been precluded by the overlap of multiple effects in frequency and time. In contrast, our interferometric technique and Fourier harmonic analysis enables to  disentangle electronic couplings, allowing us to analyze photon dressing and light-induced hybridization of Floquet states.

For even larger driving strengths, both the light-matter coupling and the photoemission process itself become non-perturbative. This largely unexplored territory of strong-field photoemission exhibits features beyond the adiabatic switching of photon dressing. Both from the experiments and the theory, we identified fracturing of the photoemission peaks in multiple-branches in the strong-field regime and attribute them to non-adiabatic LZ tunneling among the dressed states. In particular, the multi-branches structure and their relative phase shifts reveal the details of the non-adiabatic excitation. Thus the interferometric measurements allows us to trace the the dynamics of each individual band in detail on the natural time scale of the photon dressing.

Illuminating light-matter dressing in real time has profound implications for the potential of Floquet engineering in materials. In particular, the build-up and destruction of the Floquet states depend on the detailed control over the driving, thermalization and decoherence effects~\cite{aeschlimann_survival_2021,schuler_how_2020,heide_electronic_2021,ito_build-up_2023}. Disentangling the coherent dynamics from the competing decoherence processes is crucial for designing regimes that enable Floquet engineering.

On more general grounds, interferometric multi-photon photoemission provides analysis of mechanisms behind the non-pertubative strong-field phenomena~\cite{tritschler_extreme_2003,mucke_carrier-wave_2004,richter_strong-field_2024} such as HHG or THz lightwave dynamics. The LZ picture and its generalization to Floquet states is at the heart of strongly-driven dynamics. For instance, in HHG from solids~\cite{ghimire_observation_2011,vampa_linking_2015}, the interplay between complex band dispersion or topology~\cite{schmid_tunable_2021,cheng_efficient_2020} often lead to non-perturbative and nonlinear spectra at higher order. In particular, the nonlinear dynamic Bloch oscillation in Dirac cones gives rise to efficient HHG~\cite{hafez_extremely_2018,yoshikawa_high-harmonic_2017}. The LZ mechanism~\cite{wu_multilevel_2016,stehlik_role_2016} and its connection to quantum geometry of the Bloch electrons~\cite{baykusheva_strong-field_2021,schmid_tunable_2021} play key roles in HHG.
Furthermore, the lightwave currents in strongly-driven graphene can be understood with the Landau-Zener-St\"uckelberg-Majorana interference from non-adiabatic tunneling~\cite{higuchi_light-field-driven_2017}. Such transient strong field phenomena~\cite{rodionov_floquet_2016,mciver_light-induced_2020} and the underlying rich real time dynamics can be naturally described in the language of IFS.
%\textcolor{red}{In addition, we also observe chaotic like behavior in the wavefunction evolution with IFS, which comes from the subtle difference in the LZ transition matrix and details of the avoided crossing gaps (See SM for more discussion).}

A direct view into the microscopic processes underlying these strong-field phenomena has remained elusive, as the inherent momentum- and time integration of optical or transport experiments mask the non-adiabatic transitions. Our work paves a way for an in-depth understanding of strong field dynamical processes and can benefit future development of ultrafast lightwave electronics.

\section*{Acknowledgment}

We would like to thank Michael A. Sentef for fruitful discussion in the early stage of the project. M.R. acknowledges funding by the Deutsche Forschungsgemeinschaft (DFG, German Research Foundation) via 217133147/SFB 1073, project B10 and support through the Feodor Lynen Fellowship Program from the Alexander von Humboldt Foundation. M.S. and Y.Y. acknowledge support from SNSF Ambizione Grant No. PZ00P2-193527. H.P. acknowledges support from 28) the	NSF CHE-2102601 grant.

\section*{Methods}
\subsection*{Experiments}
The interferometrically time-resolved multi photon photoemission experiments (ITR-mPP) have been performed in the in-house laboratory for time-resolved photoemission spectroscopy at the University of Pittsburgh. The optical and the ultrahigh-vacuum setups are described in refs.~\cite{reutzel_coherent_2019,Cui14natphys}. In addition, ref.~\cite{reutzel_coherent_2020} provides a detailed discussion on data handling and Fourier analysis of the ITR-mPP experiments.

The single-crystal Cu(111) surface is prepared by repeated cycles of sputtering and annealing under ultra-high vacuum conditions ($ < 10^{-10}$~mbar). The quality of the surface is judged by the observation of sharp and dominant photoemission signatures from the SS and the IP state in static mPP experiments. 20~fs infrared laser pulses are generated with a noncolinear optical parametric amplifier (NOPA) pumped by a fiber laser oscillator-amplifier system operating at 1~MHz pulse rate (Clark MXR Impulse). The pulse duration is evaluated by interferometric autocorrelation mPP measurements on the polycrystalline tantalum sample holder. The vacuum electric field strength F$^{\rm exc}_0$ is on the order of 0.1-4~V/m and can be adjusted by tuning the distance between the focusing lens and the sample while keeping the average power constant. The phase-locked pulse-pair for the ITR-mPP excitation emerges from  a passively stable, self-built Mach-Zehnder interferometer~\cite{Petek97pss}. The $p$-polarized pulse-pair light illuminates the sample at an angle-of-incidence of 45°. Photoelectron energy-momentum spectra are recorded at each 100~as increment of pump-probe time-delay with a hemispherical energy analyzer (SPECS Phoibos 100).

\subsection*{Floquet quasienergy calculation}
For a time periodic Hamiltonian $\hat{H}(t+T) = \hat{H}(t)$ with frequency $\omega = 2\pi / T $, the solutions to time-dependent Schr{\"o}dinger equation (TDSE) are Floquet states $|\psi_{\alpha}(t) \rangle$, which are labeled with quasienergy $\epsilon_{\alpha}$ as
    \begin{equation}
        |\psi_{\alpha}(t)\rangle = e^{-i\epsilon_{\alpha} t} |\Phi_{\alpha}(t)\rangle = e^{-i\epsilon_{\alpha} t} \sum_m e^{-im\omega t} |\phi^{m}_{\alpha}\rangle \ ,
    \end{equation}
    where $| \phi_{\alpha}^{m} \rangle$ is the m$^{th}$ harmonic of the time periodic orbital $| \Phi_{\alpha}(t) \rangle= | \Phi_{\alpha}(t+T) \rangle$. 

    We use the time-dependent Hamiltonian Eq.\eqref{eq:tsurff_ham} and 
    projected onto the subspace spanned by the basis states $i,j =$ SS, IP1. The pulse $A(t)$ is replaced with a continuous sine wave $A_{c}(t)$ to restore time periodicity as
    \begin{align}
        h^{\mathrm{sub}}(t)&=\begin{pmatrix}
            E_{\mathrm{SS}} & O_{\mathrm{SS},\mathrm{IP1}}(t) \\
            O_{\mathrm{IP1},\mathrm{SS}}(t) & E_{\mathrm{IP1}} \\
        \end{pmatrix} \ , 
        \label{eq:ham_sub}
    \end{align}
     with the off-diagonal matrix elements
    \begin{align}
        O_{ij}(t) &= A_{c}(t) \langle i|\hat{p} | j \rangle \ .
        \label{eq:off_me}
    \end{align}
    $E_{i}$ is the energy of state $i=$ SS, IP1 solved from the one dimensional model Hamiltonian with Chulkov potential (see supplementary material~\cite{supplement}). The Floquet quasienergy $\epsilon_{\alpha}$ can be solved from the time-independent Floquet Hamiltonian as 
    \begin{equation}
        \label{eq:floquet_ham}
        H_{mn} = \frac{1}{T}\int_{0}^{T} dt \,h^{\mathrm{sub}}(t) e^{i(m-n)\omega t}
    \end{equation}
    \begin{equation}
        \label{eq:floquet_eig}
        \sum_m (H_{mn} - m\omega\delta_{mn}) |\phi^{m}_{\alpha}\rangle = \epsilon_\alpha |\phi^{n}_{\alpha}\rangle \ .
    \end{equation}

    We use a sufficiently large index of Floquet Brillouin zone $m,n=-12...+12$, for the Floquet quasienergies  to converge. In Fig.\ref{fig:tsurff_exp}\textbf{a}, the black dots are the calculated Floquet quasienergies as a function of field strength.

\subsection*{Non-equilibrium Green's function simulation}
We build upon the embedding time-resolved photoemission theory from ref.~\cite{schuler_theory_2021} and extended it  to the non-perturbative regime. To this end, we construct the extended Hamiltonian $\hat{H}^{\mathrm{PE}}$ including SS and IP1 to a set of photoelectron final states with momentum $p$:
\begin{align}
    \label{eq:ham_subspace}
    \hat{H}^{\mathrm{PE}}(t) = &\hat{H}^{\mathrm{sub}}(t) + \sum_{p} \epsilon_p(t) \hat{d}^\dagger_p \hat{d}_p \nonumber \\ &-[ qA(t) \sum_{i, p} M_{i}(E) \hat{d}^\dagger_p \hat{c}_{i} + h.c. ] \ .
\end{align}
Here, $\hat{H}^{\mathrm{sub}}(t)=\sum_{ij} h^{\mathrm{sub}}_{ij}(t)\hat{c}_{i}^{\dagger} \hat{c}_{j}$ denotes the minimal-coupling  Hamiltonian with $h^{\mathrm{sub}}_{ij}(t)$ defined in Eq.~\eqref{eq:ham_sub}. The operators  $\hat{c}_{i}^{\dagger}$ ($\hat{c}_{i}$) denote the creation (annihilation) operator for state $i$=SS, IP1, while $\hat{d}^\dagger_p$ ($\hat{d}_p$) stand for a photoelectron state with out-of-plane momentum $p$ and energy $E=p^2/2$. The vector $A(t)$ describes the Gaussian profile of the pulse. The dressing of the photoelectrons in the laser field is incorporated by the time-dependent energy $\epsilon_p(t) = (p - q A(t))^2 / 2$. In what follows we assume the continuum photoelectrons has been discretized and use the final state energy $E$ as a discrete quantum number.
The photoemission matrix elements are calculated in the velocity gauge as $M_{i}(E) = \langle \chi_E |\hat{p}| i \rangle$, where $|\chi_E\rangle$ denotes the scattering state with outgoing boundary conditions calculated from the Chulkov potential (see supplementary material~\cite{supplement} for details).

We define the one-particle Green's function in the subspace of surface states by $G_{ij}(t,t')=-i \langle T_C \hat{c}_i(t) \hat{c}^\dagger_j(t') \rangle$. The Green's function is obtained as the solution to the Kadanoff-Baym equation (KBE):

\begin{align}
    \label{eq:kbe}
    \left[i\frac{d}{dt} - \mathbf{h}^{\mathrm{sub}}(t)\right]\mathbf{G}(t,t') = \delta_C(t,t') + \int_C d\bar{t}\, \mathbf{\Sigma} (t,\bar{t}) \mathbf{G}(\bar{t}, t') \ ,
\end{align}
where the bold symbol $\mathbf{G}$ denotes two-by-two matrix-valued Green's function. The matrix elements of embedding self energy $\boldsymbol{\Sigma}$ are calculated as 
\begin{align}
    \Sigma_{ij}(t,t') \equiv \sum_{E} \Sigma_{ij}(E; t, t') = \sum_E M_{i}^{*}(E) g_E(t,t') M_{j}(E) \ ,
\end{align}
where $g_E(t,t')$ is the Volkov Green's function. Finally, the current flowing into the photoelectron subsystem $\dot{N}_E(t)$ is computed from the transient Meir-Wingreen formula~\cite{stefanucci_nonequilibrium_2013}. The photoelectron intensity is then given by $I(E) = \int^\infty_0 dt\, \dot{N}_E(t)$. We thus obtain the following expression:
\begin{align}
    \label{eq:meier-wingreen}
    I(E) &\propto  \mathrm{Re} \mathrm{Tr} \int_0^{\infty} dt \int^t_0 dt'\,  \boldsymbol{\Sigma}^{\mathrm{R}}(E;t,t') \mathbf{G}^{<}(t',t)\ .
\end{align}
For the simulation in Fig.~\ref{fig:tsurff_exp} and Fig.~\ref{fig:tsurff_exp_2w}, we use a work function of 4.5 eV, and the energy axis is referenced to the Fermi energy.

\subsection*{Instantaneous Floquet states (IFS) dynamics}

Within IFS theory, time dependence is introduced to Floquet states by following the pulse profile as a function of real time ($t_r$) and constructing instantaneous Floquet states.  At time $t_r$, the Floquet states with respect to instantaneous photon energy $\omega(t_{r})$ and the pulse envelope $F_0(t_{r})$ form a complete set (Note that in the main text we use $F_0^{\mathrm{sim}}(t_{r})$ to clarify this is a field strength for simulation). An IFS $\alpha$ at real time $t_{r}$ is defined as

\begin{align}
    \label{eq:ifs}
    |\psi_{\alpha}(F_0(t_{r}),\omega(t_{r}),&t)\rangle = e^{-i\epsilon_{\alpha}(t_{r})t} \nonumber \\
    & \times \sum_m e^{-im\omega(t_{r}) t} |\phi_{\alpha,m}(t_{r})\rangle \ ,
\end{align}
where $|\phi_{\alpha,m}(t_{r})\rangle$ denotes m$^{\mathrm{th}}$ instantaneous Floquet mode and $\epsilon_{\alpha}(t_{r})+m\omega(t_{r})$ is the instantaneous Floquet quasienergy displaced into the $m^{\mathrm{th}}$ Floquet Brillouin zone. The photoemission process can be modeled as a driven three states system with SS, IP1, and a final photoelectron state. Here we focus on the dynamics within 
the first Floquet Brillouin zone above vacuum (BZ5 in Fig.~\ref{fig:schem_exp}\textbf{a} and \textbf{b}), and 
the dynamics are categorized into two types based on IFS quasienergy spectrum.    

\subsubsection*{Type \textit{I}. Adiabatic evolution for final state} When the final state energy is far away from the Floquet replicas of SS and IP1 in BZ5 ($E_{\mathrm{SS}}+4\omega_p = E_{\mathrm{IP1}}+1\omega_p$), dynamics of the final state is mostly adiabatic.  At the beginning and end of the pulse, the resonant condition between SS and IP1 makes IFS1/IFS2 degenerate. We consider a finite space-time evolution with light-matter interaction. A driven state $|\Psi(t_r) \rangle = \sum_{i \in \{\mathrm{SS},\mathrm{IP1},\mathrm{final}\}} c_{i}(t_r) |i \rangle$ follows TDSE as
    \begin{align}
            i\frac{d}{dt_r}c_{i}(t_r) &= \sum_{j} H_{ij}(t_r)c_{j}(t_r) \ ,
    \end{align}
    with the Hamiltonian
    \begin{equation}
        \label{eq:ifs_ham}
        H(t_r) = 
        \begin{pmatrix}
            E_{\mathrm{SS}}   &   O_{\mathrm{SS},\mathrm{IP1}}(t_r)   &   O_{\mathrm{SS},\mathrm{final}}(t_r) \\
            O_{\mathrm{IP1},\mathrm{SS}}(t_r)   &   E_{\mathrm{IP1}}  &   O_{\mathrm{IP1},\mathrm{final}}(t_r) \\
            O_{\mathrm{final},\mathrm{SS}}(t_r)    &   O_{\mathrm{final},\mathrm{IP1}}(t_r)   &   E_{\mathrm{final}} 
         \end{pmatrix} \ .
    \end{equation}
    The final state energy $E_{\mathrm{final}}$ and the corresponding matrix elements are solved from Chulkov model Hamiltonian with suitable boundary condition (See supplementary material~\cite{supplement}). The off-diagonal matrix elements $O_{ij}(t)$ are defined as in Eq.~\eqref{eq:off_me}. 
        
   Similarly, the driven state $|\Psi(t_r)\rangle$ can also be expanded with IFS basis in Eq.~\eqref{eq:ifs} as
        \begin{equation}
            \label{eq:driven_ifs_expand}
            |\Psi(t_r)\rangle = \sum_{\alpha m} c_{\alpha m}(t_r) |\phi_{\alpha,m}(F_0(t_r),\omega(t_r))\rangle \ .
        \end{equation}
    Due to the energy-periodicity of the Floquet states and unitarity, we have $c_{\alpha m}(t_r) = \sum_i V_{\alpha i}(t_r) c_{i}(t_r)/\sqrt{N_f}$, where $N_f=25$ denotes the truncated number of Floquet Brillouin zone and we approximate the transformation between real states ($i=$SS, IP1, final) and the IFS ($\alpha=$IFS1, IFS2, IFS3) as a three-by-three identity matrix $\mathbf{V}(t_r)=\mathbf{I}$ due to the adiabaticity. Such approximation is justified by the fact that the character of each IFS does not evolve significantly during type \textit{I} dynamics, as shown in Fig.~\ref{fig:ifs}\textbf{b} and \textbf{e}. In a Floquet Brillouin zone $m_0$, we denote $c_{\alpha m_0}(t_r)$ as a column vector $\mathbf{c}_{m_0}(t_r)$. The evolution between time $t_1$ and $t_2$ can then be expressed as a unitary evolution $\mathbf{\hat{U}}_{m_0}(t_2,t_1)\mathbf{c}_{m_0}(t_1) = \mathbf{c}_{m_0}(t_2)$, following the Hamiltonian in Eq.~\eqref{eq:ifs_ham}.

    \subsubsection*{Type \textit{II}. Landau-Zener mechanism among IFS}
    When the final state energy is close to the Floquet replicas of the surface states in BZ5 ($E_{\mathrm{SS}}+4\omega_p = E_{\mathrm{IP1}}+1\omega_p$), avoided crossings among IFS are formed and the dynamics becomes non-adiabatic. In the vicinity of the avoided-crossing points (AC1/AC2) formed by the IFS $(\alpha,m)$ and $(\alpha',m')$, the population transfer can be described by a Landau-Zener transfer matrix in Eq.~\eqref{eq:lz_matrix}. The parameters related to the details of the avoided-crossing are defined as 
    \begin{align}
        \label{eq:lz_parameters}
        \delta_{\alpha m \alpha' m'} &= \frac{\Delta_{\alpha m \alpha' m'}^2}{4} \ , \nonumber \\
        P_{\alpha m \alpha' m'} &= exp(-2\pi \delta_{\alpha m \alpha' m'}) \ , \nonumber \\
        \phi_{\alpha m \alpha' m'}^{s} &= \frac{\pi}{4}+ \delta_{\alpha m \alpha' m'} ln(\delta_{\alpha m \alpha' m'}-1) \nonumber \\ &+ arg \Gamma (1-i\delta_{\alpha m \alpha' m'}) \ .
    \end{align}
    
     Here, $\Delta_{\alpha m \alpha' m'}$ and $v_{\alpha m \alpha' m'}$ are the gap and passing speed of the avoided crossings between IFS ($\alpha$,m) and ($\alpha'$,$m'$). $\phi^s$ denotes the Stokes phase, and $\Gamma(z)$ is the gamma function.

    In Fig.~\ref{fig:ifs}\textbf{e}, the dynamics is seperated into \ I $\rightarrow$ II $\rightarrow$ I $\rightarrow$ II $\rightarrow$ I based on instantaneous quasienergies. All the relevant avoided crossing points are formed by the IFS in the same Floquet Brillouin zone ($m_0=$BZ5 as defined in Fig.~\ref{fig:schem_exp}).The time evolution of $\mathbf{c}_{m_0}(t)$ vectors is then calculated as 
    \begin{align*}
        \label{eq:IFS_evol}
        &\mathbf{c}_{m_0}(t_r = t_{fin})  \\
        &= [\mathbf{\hat{U}}_{m_0}(t_{\mathrm{fin}}, t_{\mathrm{AC2}})\hat{\mathbf{T}}^{\mathrm{AC2}}_{\mathrm{IP1},m_0,\mathrm{TL},m_0}\mathbf{\hat{U}}_{m_0}(t_{\mathrm{AC2}},t_{\mathrm{AC1}}) \\ &\times \hat{\mathbf{T}}^{\mathrm{AC1}}_{\mathrm{IP1},m_0,\mathrm{TL},m_0}\mathbf{\hat{U}}_{m_0}(t_{\mathrm{AC1}},t_{\mathrm{ini}})] \mathbf{c}_{m_0}(t_r = t_{\mathrm{ini}}) \ ,
    \end{align*}
    where $t_{\mathrm{ini}}$ and $t_{\mathrm{fin}}$ denote the starting and the ending time of the pump+probe pulse. At $t=t_{ini}$, only SS is occupied, so we set $c_{\mathrm{IFS1},m_0}(t_r=t_{\mathrm{ini}})=1/\sqrt{N_f}$. The decoherence effect is not considered in our IFS simulation, which would give an effective energy broadening, because the decoherence time would be much longer than the pulse duration \cite{reutzel_coherent_2020}. Therefore we employ an effective energy Gaussian smearing in the time delay and $E_{\mathrm{final}}$ scan in Fig. ~\ref{fig:ifs}\textbf{d}-\textbf{i}, to generate a similar broadening as observed in the experiment.

\bibliography{mPP-floquet_paper_11}

%TC:endignore

%\clearpage

%\section*{Competing Interest}
%The authors declare no competing interest.

%\section*{Data availability}
%The data of this study will be made available on the Open Research Data Repository of the Max Planck Society. (reserved DOI: \href{https://doi.org/10.17617/3.PILCPQ}{10.17617/3.PILCPQ} )

%\section*{Methods}
%\subsection*{Time dependent surface flux method (t-SURFF)}

\end{document}

%\author{authors}
%\affiliation{affiliations}

\author{Yun~Yen}
\affiliation{PSI Center for Scientific Computing, Theory, and Data, 
5232 Villigen PSI, Switzerland}
\affiliation{École Polytechnique Fédérale de Lausanne (EPFL), Switzerland}

\author{Marcel Reutzel}
\affiliation{I. Physikalisches Institut, Georg-August-Universit\"at G\"ottingen, G\"ottingen, Germany}

\author{Andi Li}
\affiliation{Department of Physics and Astronomy and Pittsburgh Quantum Institute, University of Pittsburgh, Pittsburgh, Pennsylvania, USA}

\author{Zehua Wang}
\affiliation{Department of Physics and Astronomy and Pittsburgh Quantum Institute, University of Pittsburgh, Pittsburgh, Pennsylvania, USA}

\author{Hrvoje Petek}
\affiliation{Department of Physics and Astronomy and Pittsburgh Quantum Institute, University of Pittsburgh, Pittsburgh, Pennsylvania, USA}
\affiliation{Department of Physics and Astronomy and IQ Initiative, University of Pittsburgh, Pittsburgh, Pennsylvania, USA}

\author{Michael~Sch\"uler}
\email{michael.schueler@psi.ch}
\affiliation{PSI Center for Scientific Computing, Theory, and Data, 
5232 Villigen PSI, Switzerland}
\affiliation{Department of Physics, University of Fribourg, CH-1700 Fribourg, Switzerland}

\title {Supplementary Material : Observation of non-adiabatic Landau-Zener tunneling among Floquet states}

%\author{authors}

\maketitle

\makeatletter
\renewcommand{\theequation}{S\arabic{equation}}
\renewcommand{\thefigure}{S\the\numexpr\value{figure}}
\renewcommand{\bibnumfmt}[1]{[S#1]}
\renewcommand{\citenumfont}[1]{S#1}

\def\tagform@#1{\maketag@@@{(\ignorespaces{#1}\unskip\@@italiccorr)}}
\renewcommand{\eqref}[1]{\textup{{\normalfont(\ref{#1}}\normalfont)}}
\makeatother

\renewcommand{\figurename}{Figure}
\renewcommand{\thesection}{Supplementary~Note~\arabic{section}}

\tableofcontents

% \clearpage
\section{1D surface Chulkov model Hamiltonian}
    \label{section:1d_model}
 To describe the image charge nature of the IP1 state close to the metal surfaces, we apply Chulkov potential \cite{chulkov_image_1999}, which gives a suitable mean-field correction to reproduce accurate band gaps between the surface states and image potential states. Here we modify the expression to take vacuum into account, and fit the parameters based on Cu(111) surface band gaps at $\Gamma$ point. The modified Chulkov potential $V_c(z) = \sum_{i=1}^{6} V_i(z)$ reads as

    \begin{align}
        \label{eq:chulkov1}
        V_1(z) &= A_{10} + A_{1} cos(\frac{2 \pi}{a_s}z) \ , z<D \ , \\
        \label{eq:chulkov2}
        V_2(z) &= -A_{20} + A_{2} cos(\beta (z-D))\ , D<z<z_1 \ ,\\
        \label{eq:chulkov3}
        V_3(z) &= A_{3} exp[-\alpha (z-z_1)] \ , z_1<z<z_{im} \ ,\\
        \label{eq:chulkov4}
        V_4(z) &= -\frac{exp(-\lambda (z-z_{im}))-1}{4(z-z_{im})} \ , z_{im}<z<z_v \ ,\\
        \label{eq:chulkov5}
        V_5(z) &= b_0 (z-z_0)^2, z_{v}<z<z_0 \ ,\\
        \label{eq:chulkov6}
        V_6(z) &= 0, z_0<z \ .
    \end{align}

\begin{figure*}[t]
        \includegraphics[width=1.0\textwidth]{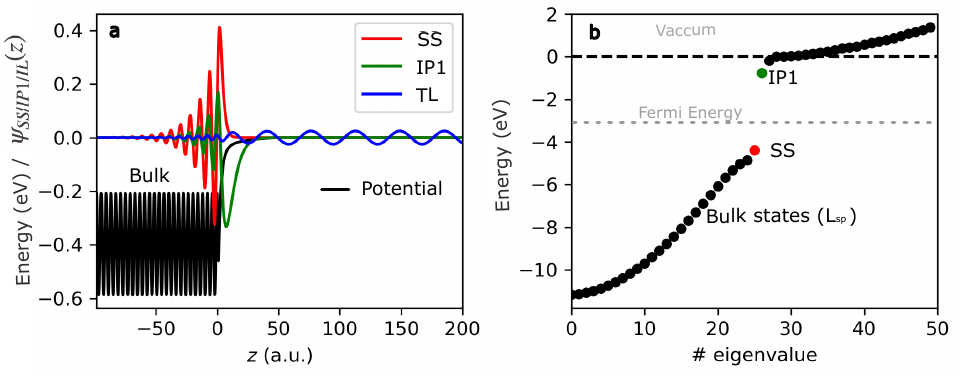}
        \caption{\textbf{Modified Chulkov potential and the surface states.} \textbf{a} The modified potential and wavefunction for Shockley surface state (SS), the first image potential state (IP1), and an exmplified time-reversed LEED state (TL) with energy 5.4 eV above vacuum, plotted as a function of the 1D real-space grid with atomic unit (a.u.). \textbf{b} Eigenenergy spectrum solved from the modified potential. Each dot represent one eigenenergy.}
    \label{fig:surface}
\end{figure*}

    Here, $V_1(z)$ describes the periodic bulk potential. The image potential term $V_4(z)$ describes the coulomb potential between an external charge and its image charge, which gives rise to the image potential states. The original Chulkov potential only contains $V_1(z)$ to $V_4(z)$, and we define $V_5(z)$ so the surface potential can be continuously connected to the vacuum $V_6(z)=0$, in order to solve for photoelectron final states. The black line in Fig.~\ref{fig:surface}\textbf{a} shows $V_c(z)$ with the parameters we choose.

    The full Hamiltonian can then be expressed using a 1D real space discretized basis $\{ \ket{z_i} $,$i=1..N$\}. The Hamiltonian matrix element is defined as $H_{ii'}=\langle z_i|\hat{H}|z_{i'} \rangle = (-1/2)K_{ii'} + V_c(z_i) \delta_{ii'}$, where $K_{ii'}$ denotes the second order derivative on the discretized grid. Here we use the first-order expression as $K_{ii}=-2/\Delta z^2$,  $K_{i,i+1}=K_{i,i-1}=1/\Delta z^2$ with a uniform grid spacing $\Delta z=z_{i+1}-z_i$. By fitting the eigenenergy spectrum in Fig.~\ref{fig:surface}\textbf{b} to the experimental band gaps~\cite{chulkov_image_1999}, in particular the gap between first image potential state(IP1) and the second image potential state (IP2) and the gap between Shockley surface state (SS) and lower bulk sp-band ($L_{sp}$), we get the optimal parameters : lattice constant $a_0 = 3.94 ${\AA} , $A_1$ = 5.14 eV, $A_2$ = 3.43 eV, $\beta = 2.94$, $z_1 = 4.71 \pi / (4\beta)$, $z_v = 20 ${\AA}. $z_0$ and $\alpha$ can be deduced from continuity of the potential. Fig.\ref{fig:surface}\textbf{a} also shows the wavefunction for SS (red) and IP1 (blue). Both states are well-localized near the surface, and part of the IP1 wavefunction extends outside the surface, which is consistent with its image-charge nature. For the non-equilibrium Green's function (NEGF) simulation and instantaneous Floquet states (IFS) simulation in the main text, we use the surface states energy and velocity matrix elements solved from this 1D model Hamiltonian.

    % A1 = 5.14
    % A2 = 3.43103
    % beta = 2.9416
    % z1 = 4.706897pi/4beta
    % zv = 20

    \subsection*{Time-reversed low-energy electron diffraction (TL) states}

    In the IFS simulation, we use the time-reversed low-energy electron diffraction (TL) states as phtotoelectron final states. A TL state $\chi_E(z)$ with energy $E$ obeys the asymptotic behavior in the vaccum close to the right boundary as 

    \begin{align}
            \label{eq:asymp}
            \chi_E(z) = e^{ikz}+re^{-ikz}, 
    \end{align}
    where $k=\sqrt{2E}$ is the wave number and $r$ is the reflection coefficient. We can solve the TL states on the 1D grid with the secular equation 
    \begin{align}
            \label{eq:chi_secular}
            \sum_{i'} [\Tilde{K}_{ii'} + (V_c(z_{i'})-E)\delta_{ii'}] \chi_E(z_{i'}) = B_i \, 
    \end{align}
    where $\Tilde{K}_{ii'}$ denotes the modified second order derivative with modified matrix element $\Tilde{K}_{11}=\Tilde{K}_{nn}=e^{-ik\Delta z}$. $B_i$ follows the boundary condition in Eq.~\eqref{eq:asymp}, and the only finite term is $B_{N}=e^{i k z_{N}}sin(k \Delta z)$. The kinectic energy $E$ can be selected from the continuum. An TL state with $E=5.4 eV$ is shown in Fig.~\ref{fig:surface}\textbf{a} as an example.

    \section{Explanation of the $\pi$ phase shift in strong field ITR-4PP}
    \label{section:pi_shift}
    As discussed in the main text, the central $\pi$ phase shift of ITR-4PP at strong field comes from the  \textit{triangular shape} of the finite ITR-4PP signal, which is defined by the Floquet quasienergies driven from SS and IP1. In this supplementary note, we provide three additional examples of IFS dynamics on the $E_{\mathrm{final}}$ -time delay plane to offer a clearer illustration on the formation of such ITR-4PP.
    
\begin{figure*}[t]
    \includegraphics[width=0.9\textwidth]{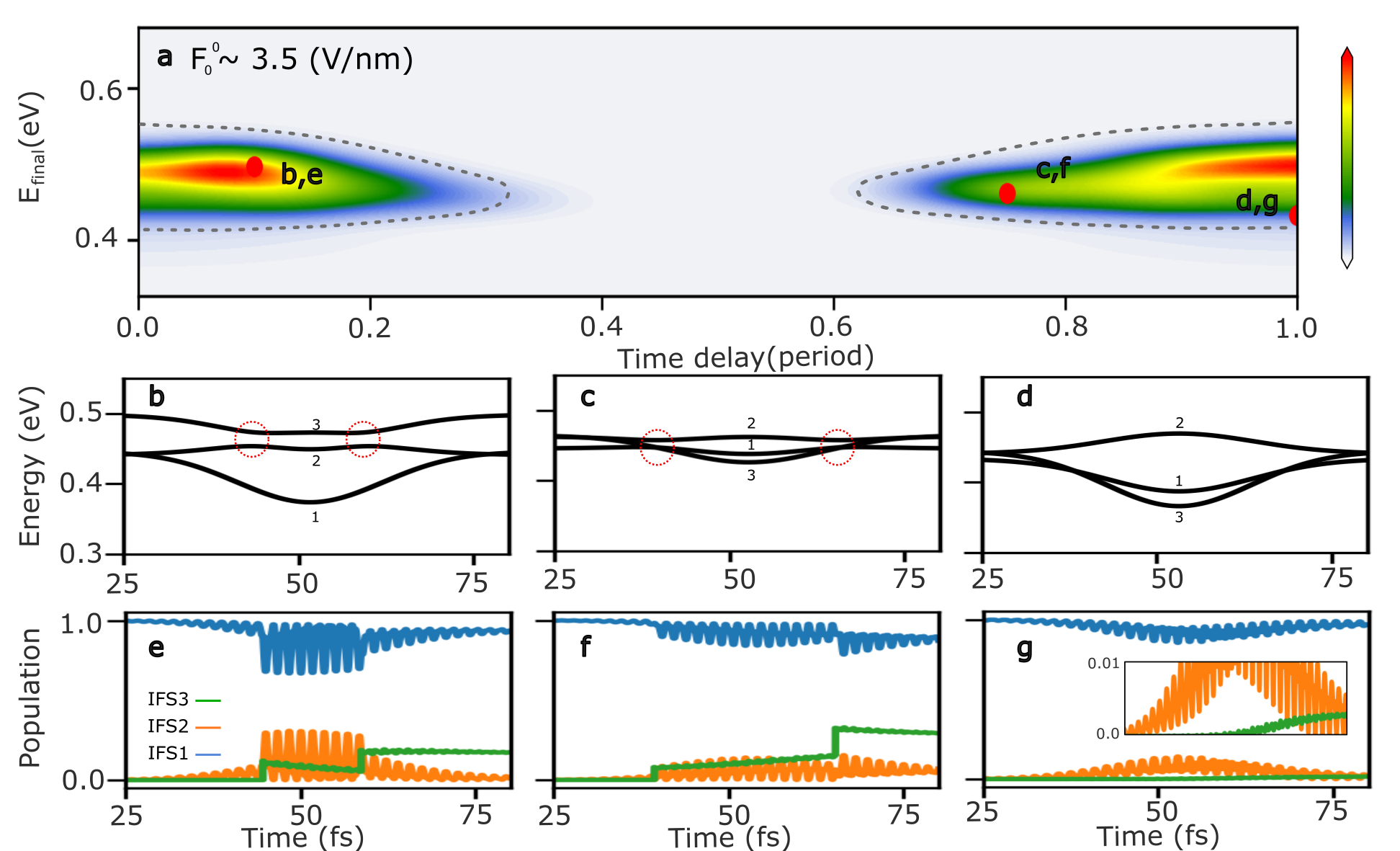}
    \caption{\textbf{Strong field IFS simulation examples} \textbf{a}  Population of IFS3 at the end of the pulse with varying final state energy ($E_{\mathrm{final}}$) and time delay for $F_0^{\mathrm{sim}} \sim 3.5$ V/nm. This is identical to fig.4\textbf{j} of the main text. \textbf{b}-\textbf{d} IFS quasienergy spectrum for three red dots in \textbf{a}. The labels 1,2,and 3 refer to IFS1, IFS2, and IFS3, respectively. \textbf{e}-\textbf{g} IFS population dynamics corresponding to \textbf{b}-\textbf{d}. The energy axis is referenced to the Fermi energy with a theoretical work function of 4.5 eV. Here we focus on the dynamics in the first Floquet Brillouin zone above vacuum (BZ5 in Fig.1 of the main text). } 
    \label{fig:ifs_example}
    \end{figure*}

    In Fig.~\ref{fig:ifs_example}\textbf{a}, we show again the population of IFS3 at $F_0^{\mathrm{sim}} \sim 3.5$ V/nm, which is identical to Fig. 4\textbf{i} of the main text. The \textit{triangular shaped} area with finite 4PP signal is highlighted with the dashed boundary lines as a guide to the eye. 
    
    In Fig.~\ref{fig:ifs_example}\textbf{b} and \textbf{e}, two pulses are delayed only by 0.1 period and $E_{\mathrm{final}}\approx $ 5.0 eV, giving a strong effective field strength and large energy dressing between IFS1 and IFS2. IFS3 and IFS2 form two avoided crossing points during the process, leading to finite population of IFS3 by Landau-Zener mechanism in the end.  In Fig.~\ref{fig:ifs_example}\textbf{d} and \textbf{g}, two pulses are delayed by a full period and therefore they are completely in phase. A lower $E_{\mathrm{final}}\approx $ 4.93 eV is chosen, at which IFS3 \textit{"touches"} IFS2 without forming avoided crossing gaps. Indeed the small population of IFS3 mainly comes from the coherent oscillation as shown in Fig.~\ref{fig:ifs_example}\textbf{g}. This point is at the "lower boundary" of the finite IFS3 population region. In fact, the entire "lower boundary line" of the finite population region is defined by the parameter space where IFS3 touches IFS1. Similarly the "upper boundary line" corresponds to where IFS3 touches IFS2. As time delay gets closer to 0.5 period, where the sum of two pulses is effectively weak, the splitting between IFS1/IFS2 becomes smaller, leading to a narrower range of finite IFS population along the  $E_{\mathrm{final}}$ axis. This explains why the finite population area has the \textit{triangular shape} in Fig.~\ref{fig:ifs_example}\textbf{a}. 
    
    If the peak field is sufficiently strong, the finite population area extends closer to the middle of the plot (half cycle time delay) - the triangular shape elongates toward time delay $\sim$ 0.5 period in Fig.~\ref{fig:ifs_example}\textbf{a}. For $F_0^{\mathrm{sim}}\sim 3.5$ V/nm, some photoemission signal remains finite at time delay $\sim 0.75$ period and $\sim 0.25$ period around $E_{\mathrm{final}} \sim 0.5$ eV, as demonstrated in Fig.~\ref{fig:ifs_example}\textbf{c} and \textbf{f}. This naturally gives rise to the $\pi$ phase shift in $2\omega_p$ harmonics at $E_{\mathrm{final}} \sim 0.5$ eV. This argument is supported by our NEGF simulation, where the Floquet quasienergies clearly define the boundary of ITR-4PP in Fig.~\ref{fig:exp_lape}\textbf{a}-\textbf{d}.

\section{Laser-assisted photoemission effect (LAPE) and 5PP}
    \label{section:lape_ifs}
\begin{figure*}[t]
    \includegraphics[width=0.9\textwidth]{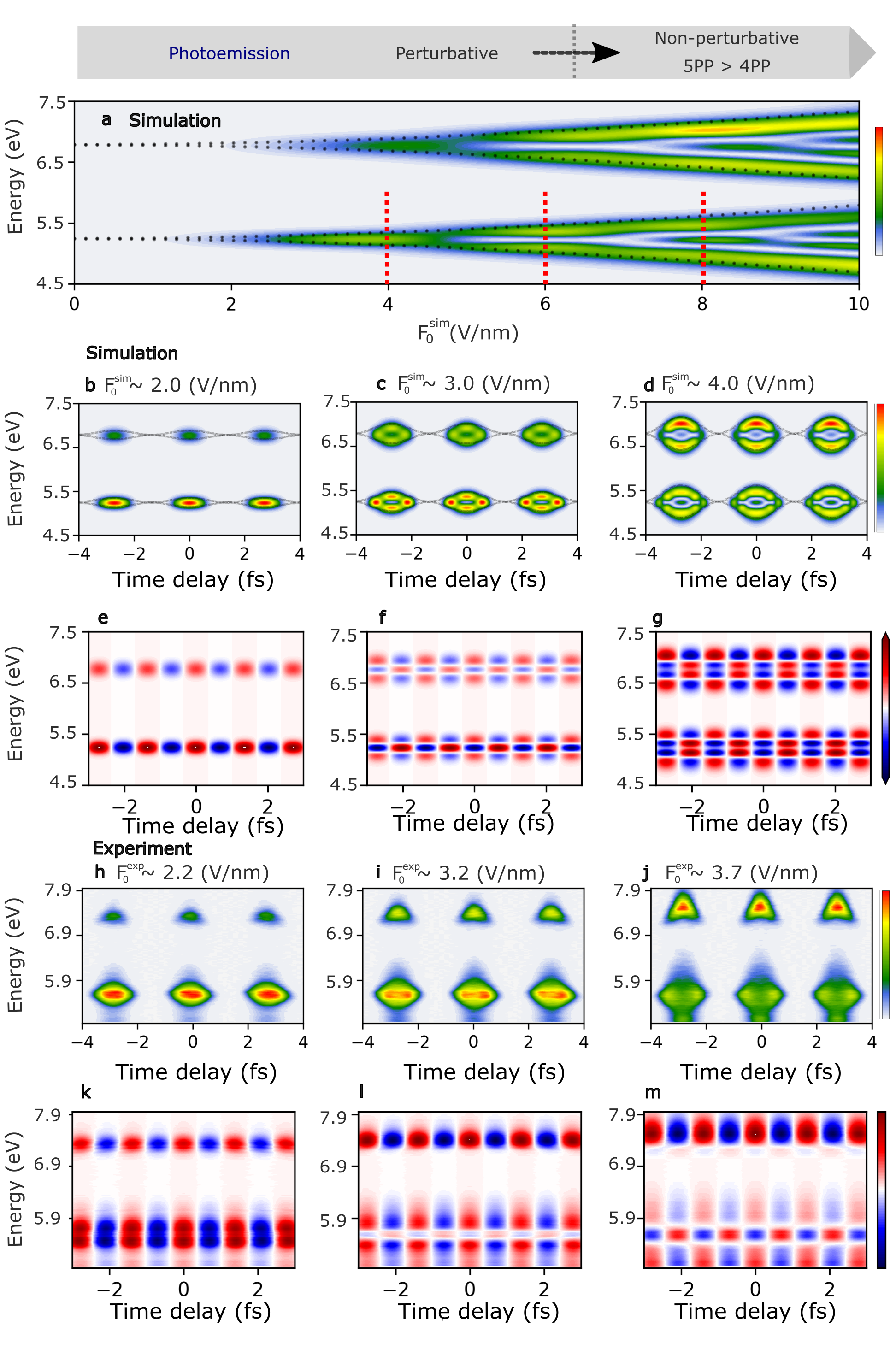}
    \caption{ \textbf{Comparison of 4PP and 5PP for simulation and experiments} \textbf{a} NEGF simulation with Floquet quasienergy plotted in the dashed line. \textbf{b}-\textbf{d} ITR-mPP simulations for 4PP and 5PP with theoretical field strength $F_0^{\mathrm{sim}} \sim$ 2.0, 3.0, and 4.0 V/nm. The black dots representing the corresponding Floquet quasienergies. \textbf{e}-\textbf{g} 2$\omega_p$ harmonics from the simulation in \textbf{b}-\textbf{d}. \textbf{h}-\textbf{j} Experimental ITR-4PP data with estimated field strength $F_0^{\mathrm{exp}} \sim$ 2.2, 3.2, and 3.7 V/nm. \textbf{k}-\textbf{m} Experimental 2$\omega_p$ harmonics corresponding to \textbf{h}-\textbf{j}. The energy axis is referenced to the Fermi energy with a theoretical work function of 4.5 eV.} 
    \label{fig:exp_lape}
    \end{figure*}

    As the field intensifies, the 5PP intensity becomes stronger than 4PP at some point, which marks the crossover to non-perturbative photoemission regime ~\cite{reutzel_coherent_2020}. This is demonstrated both by our NEGF simulation in Fig.~\ref{fig:exp_lape}\textbf{a}-\textbf{d} and the experimental mPP data in Fig.~\ref{fig:exp_lape}\textbf{h}-\textbf{j}, which are the same set of data as Fig.2 in the main text with an extended range of energy shown, to compare the 4PP and 5PP spectra. The central peak and the multifold splitting emerge at a field strength $F_0^{\mathrm{sim}} \sim 7$ V/nm, which is close to where the non-perturbative crossover occurs. As we argue in the~\ref{section:pi_shift} and Fig.~\ref{fig:ifs_example}, the overall shape and boundary of ITR-mPP are defined by the Floquet quasienergies. Moreover, the detailed features of ITR-mPP can be related to LAPE. 
    
    Our NEGF simulation captures LAPE and the non-perturbative photoemission to some extent, while the actual intricate surface screening profile can lead to detailed features beyond our simulation.  One obvious difference can be seen from the comparison of 5PP in strong field : The experiment in Fig.~\ref{fig:exp_lape}\textbf{j} shows only up-bending branch, while the simulation in Fig.~\ref{fig:exp_lape}\textbf{d} shows two Floquet splitting branches with asymmetric intensity. However, the boundary of 5PP in simulation is still defined by the Floquet quasienergies.

    The IFS simulation in the main text uses a three states model, which already captures strong field 4PP features. However, the simulation neglects LAPE completely. To study how LAPE affect the IFS dynamics, we include a second time reversed LEED state with energy shifted by one photon quanta compared to the first TL ($E_{\mathrm{TL2}} = E_{\mathrm{TL1}} + \omega_p$), and simulate the dynamics of these 4 IFS. Similar to the simulation of 3 IFS, both TL1 and TL2 gains population via Landau-Zener tunneling and coherent dynamics from the coupling to the surface states. The population of TL1 and TL2 can be interpreted to be proportional to 4PP and 5PP respectively.
    
\begin{figure*}[t]
    \includegraphics[width=0.9\textwidth]{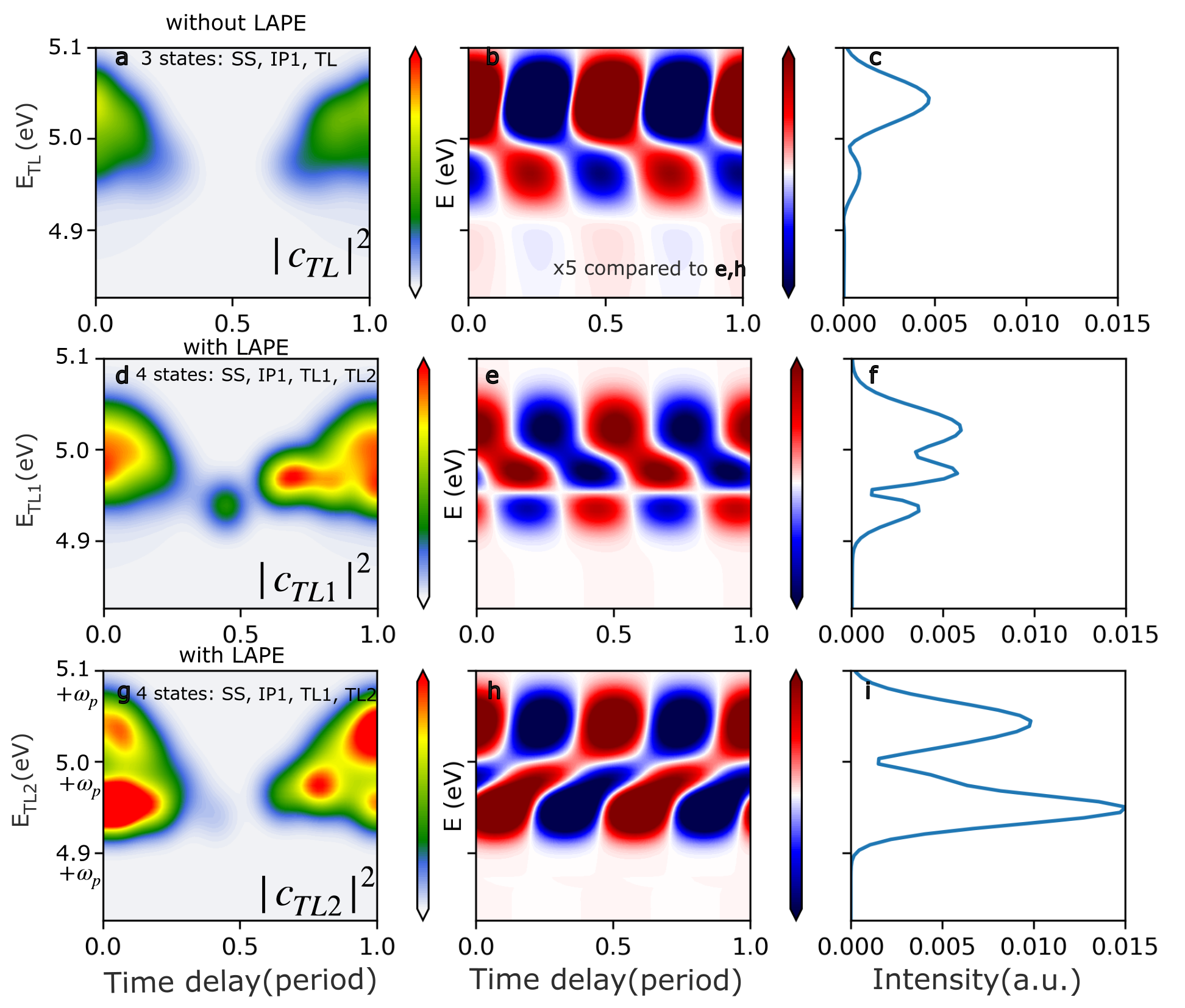}
    \caption{ \textbf{3-states and 4-states IFS simualtion for $F_0^{\mathrm{sim}} \sim 4.5$ V/nm} \textbf{a} The population of the IFS3 (TL) at the end of the pulse, \textbf{b} the corresponding 2$\omega_p$ harmonics, and \textbf{c} the integrated intensity in the 3-states calculation. The 4-states calculation includes two TL states with energy $E_{\mathrm{TL2}} = E_{\mathrm{TL1}} + \omega_p$. \textbf{d}-\textbf{f} shows the population, $2\omega_p$ harmonics, and the integrated intensity for the population of IFS3 (TL1). \textbf{g}-\textbf{i} The population, 2$\omega_p$ harmonics, and the integrated intensity for the population of IFS4 (TL2).}
    \label{fig:ifs_lape}
    \end{figure*}

    Here we take a stronger field $F_0^{\mathrm{sim}} = 4.5$ V/nm, compared to the IFS simulation in the main text ($F_0^{\mathrm{sim}} = 2.5$ and 3.5 V/nm), in order to show a clear comparison between 4PP and 5PP. In Fig.~\ref{fig:ifs_lape}\textbf{a}-\textbf{c}, the population of IFS3 (TL) from the 3-states simulation is shown, where the phase shift is present as the experimental data. Nevertheless, the top splitting branch is much stronger than the bottom splitting branch, which is similar to Fig. 4\textbf{i} and \textbf{j} in the main text. In the 4-states calculation, the additional TL2 opens a new channel for Landau-Zener tunneling. Since the matrix element between IP1-TL2 is much stronger than the matrix element between SS-TL2, the 5PP has a much stronger upper branch, as shown in Fig.~\ref{fig:ifs_lape}\textbf{g}-\textbf{i}. In fact this is consistent with the strong field ($F_0^{\mathrm{exp}} = 3.7$ V/nm) experimental data in Fig.~\ref{fig:exp_lape}\textbf{j} and \textbf{m}, where 5PP only has the shape of upper triangles. Effectively the three branches of 4PP become more symmetric in Fig.\ref{fig:ifs_lape}\textbf{d}-\textbf{f}. Such relative intensity of 4PP is closer to what we observe in the experiments and the NEGF simulation.

%%%%%%%%%%%%%%%%%%%%%%%%%%%%%%%%%%%%%%%%%%%%%%%%%%%%%%
 \FloatBarrier
% \clearpage

\bibliographystyle{apsrev4-2}
\bibliography{mPP-floquet}